\documentclass[11pt,dvips]{article}
\usepackage{graphicx}
\usepackage{amsmath}
\usepackage{amssymb}
\usepackage{latexsym}
\usepackage{color}
\begin{document}
\thispagestyle{empty}
\begin{center}

\vspace{1.8cm}


 {\Large {\bf A  polynomial class of $u(2)$ algebras}}\\

\vspace{1.5cm} {\bf Mohammed Daoud}$^{a,b}${\footnote { email: {\sf
m$_{-}$daoud@hotmail.com}}} and {\bf  Won Sang Chung }$^{c}$ {\footnote
{ email: {\sf mimip4444@hanmail.net}}}  \\
\vspace{0.5cm}
$^{a}${\it Abdus Salam  International Centre for Theoretical Physics, Trieste, Italy}\\[0.5em]
$^{b}${\it Department of Physics, Faculty of Sciences,  University Ibnou Zohr, Agadir , Morocco}\\[0.5em]
$^{c}${\it Department of Physics and Research Institute of Natural Science,\\ College of
Natural Science, Gyeongsang National University, Jinju 660-701, Korea}\\[0.5em]

\vspace{3cm} {\bf Abstract}
\end{center}
\baselineskip=18pt
\medskip

\noindent A $r$-parameter   ${u}_{\{\kappa_1, \kappa_2, \cdots, \kappa_r\}}(2)$  algebra is introduced. Finite unitary representations are
investigated. This polynomial algebra reduces via a contraction procedure to the generalized Weyl-Heisenberg algebra ${\cal A}_{\{\kappa_1, \kappa_2, \cdots, \kappa_r\}}$
(M. Daoud and M. Kibler, J. Phys. A: Math. Theor. {\bf 45} (2012)
244036).  A pair of nonlinear (quadratic) bosons of type ${\cal A}_{\kappa}\equiv {\cal A}_{\{\kappa_1=\kappa, \kappa_2=0, \cdots, \kappa_r=0\}}$ are used to construct, \`a la Schwinger, a one parameter family of (cubic) $u_{\kappa}(2)$ algebra. The corresponding
Hilbert space is constructed. The analytical Bargmann representation is also presented.


\vspace{1cm}

\newpage
\section{Introduction}

\noindent

Nonlinear extensions of finite Lie algebras continue to attract much attention both in mathematics and physics.
The special interest in nonlinear algebras is mainly motivated by their appearance as dynamical symmetries for several
physical systems (two identical particles in two dimensional manifold, isotropic oscillator and Kepler system in a two dimensional  curved space, ...)
They have also  found various  applications in other areas of physics such as quantum statistics, quantum optics, etc. [1-17].
For  bosonic realizations of such algebras, different kinds of generalized boson algebra were introduced in the literature. Different extensions give
rise to different kinds of structure relations and subsequently different types of unitary representations. One may quote for instance the
 concept deformed boson ($q$-boson) algebra, firstly proposed by Arik and Coon \cite{Arik} and developed
further by Macfarlane \cite{Macfarlane} and Biedenharn \cite{Biedenharn}, which was used
extensively in the literature to provide bosonic realizations of quantum algebras. \\

 Now, it is well established
that all generalized or deformed bosons can be accommodated within a unified framework in which
the product of creation and annihilation operators is a function of the number operator (see for instance \cite{Manko}). This structure function reflects
the nonlinearity effects and the deviation from the usual boson algebra. Of particular interest are structure functions which
expand as polynomials with respect in the number operator.  This special kind of nonlinear oscillator
algebra is commonly called in the literature polynomial Weyl-Heisenberg algebras (see \cite{daoud1,daoud2,daoud3} and references quoted therein).
In this context, recently, the $r$-parameter
 Weyl-Heisenberg algebra $ {\cal A}_{\kappa_1,\kappa_2, \cdots,\kappa_r} $ was introduced. It generalizes the single mode algebra
 $ {\cal A}_{\kappa} $  which covers the usual boson, $su(2)$, $su(1,1)$ algebras \cite{daoud1}. Also, this
 algebra turns out to be of special interest in dealing with some one dimensional quantum potentials  (for more detail see \cite{daoud1}). The representation theory
 of algebras of type $ {\cal A}_{\kappa_1,\kappa_2, \cdots,\kappa_r} $ is very rich. Indeed, it has been shown that
they admit finite and finite dimensional unitary representations depending on the values of the parameters $\kappa_1,\kappa_2, \cdots,\kappa_r$. Multi-mode
extensions of the algebra   $ {\cal A}_{\kappa} $ and the related representations were investigated in \cite{daoud2,chung1}. Furthermore, the fermionic analogue of  the ${\cal A}_{\kappa}$ algebra
  was proposed in \cite{chung2,chung3}. \\

In addition, the generalized $su(2)$  algebras developed in the context of the theory of quantum algebras have
introduced in a way that their finite dimensional unitary representations remain as close as possible to usual one. The interest
in such nonlinear algebraic structure is due to their relevance in diverse areas such as integrable models
\cite{Floreanini,Myrheim,Bonatsos2,Dask,Marquette} and super-symmetric quantum systems \cite{Marquette,Debergh}.
Similarly, to generalized boson algebras, the nonlinear  $u(2)$ algebras can be
defined within  an unified framework in which the commutator, involving the raising and lowering generators $J_+$ and $J_-$,
is expanded as formal power series of the Cartan subalgebra $J_0$ and $J_3$. Unitary irreducible representations were investigated
  for some specific forms of the polynomial structure
function $f( J_3 )\equiv [J_+ , J_-]$ (see for instance \cite{abdess}). \\

In this work, we shall focus on the polynomial extension of $u(2)$ algebra by adopting
a procedure similar to one giving the algebra $ {\cal A}_{\kappa_1,\kappa_2, \cdots,\kappa_r}$ \cite{daoud2}. In this scenario,
we introduce the $r$-parameter $u_{\kappa_1,\kappa_2, \cdots,\kappa_r}(2)$. The structure function defined by
means of the commutator $[J_+ , J_-]$ is a polynomial of order $r$ in the Cartan generators $J_0$ and $J_3$. The second
facet of this paper concerns the Schwinger realization of a polynomial $u(2)$ algebra by means of a pair of bosons of type  ${\cal A}_{\kappa}$ \cite{daoud1}. \\

This paper is organized as follows.  Section 2 deals with a $r$-parameter extension of $u(2)$ algebra. For this polynomial
variant, denoted by $u_{\kappa_1,\kappa_2, \cdots,\kappa_r}(2)$, we investigate the finite dimensional representations which
clearly will depend on the values taking by the parameters $\kappa_1,\kappa_2, \cdots,\kappa_r$. We show by a contarction procedure that the
algebra $u_{\kappa_1,\kappa_2, \cdots,\kappa_r}(2)$ reduces to the generalized Weyl-Heisenberg algebra $ {\cal A}_{\kappa_1,\kappa_2, \cdots,\kappa_r}$
introduced in \cite{daoud3}.  Section 3 is devoted to the Schwinger realization of another kind of polynomial $u(2)$ algebras. The corresponding Fock space is
explicitly constructed. Analytical representations and the associated set of coherent states are also derived.
 Concluding remarks close this paper.

\section{Polynomial ${u}_{\{\kappa\}}(2)$ algebra}

\subsection{The  algebra}
A generalized  ${u}(2)$ algebra,  defined on $\mathbb{C}$, is spanned by four
linear operators $J_{\alpha}$ ($\alpha = 0, 3, +, -$) satisfying the commutation relations
    \begin{eqnarray}
    [J_+ , J_-] =  G(J_3,J_0) \qquad
    [J_3, J_{\pm}] = \pm J_{\pm} \qquad
    [J_0, J_{\alpha}] = 0.
    \label{algebre}
    \end{eqnarray}
Since we are interested in unitary representations, we require the following hermitian conjugation conditions
    \begin{eqnarray}
    J_+ = (J_-)^{\dagger} \qquad J_0 = J_0^{\dagger} \qquad J_3 = J_3^{\dagger}.
    \label{hermiticity conditions}
    \end{eqnarray}
Thus,  the $G$ function in (\ref{algebre}) should satisfy
    \begin{eqnarray}
    G(J_3,J_0) = \big(G(J_3,J_0)\big)^{\dagger}.
    \label{hermiticity of G}
    \end{eqnarray}
 It must be  emphasized that  different polynomial extensions were considered in the literature
leading to different structures relations and consequently different types of representations (a complete list of references can be found in
\cite{Shreecharan}).  In this work, we shall
consider the $r$-parameters ${{u}}_{\{\kappa\}}(2)\equiv {u}_{\{\kappa_1, \kappa_2, \cdots, \kappa_r\}}(2)$ algebra
characterized by the structure function $\Phi(J_3,J_0)$ defined by
\begin{eqnarray}
    J_+J_-= \Phi(J_3,J_0)     \qquad  J_-J_+ = \Phi(J_3+1,J_0)
    \label{Phi(J_3,J_0}
    \end{eqnarray}
so that the the $G$ function occurring in the commutator between the Weyl generators $J_+$ and $J_-$ (see eq.(\ref{algebre})) writes
\begin{eqnarray}
    G(J_3,J_0) = \Phi(J_3,J_0) - \Phi(J_3+1,J_0).
    \label{G as difference of two F}
    \end{eqnarray}
The function $\Phi$, specifying the nonlinear scheme in this work,  is defined by
    \begin{eqnarray}
    \Phi(J_3,J_0)  = [J_0+J_3] [ I + J_0-J_3]
             [ I + \kappa_1(J_0+J_3-I)] [ I + \kappa_2(J_0+J_3-I)]\cdots
             [ I + \kappa_r(J_0+J_3-I)]
    \label{generalformofF}
    \end{eqnarray}
where the $\kappa_i$'s ($i = 1, 2, \cdots , r$) are real parameters and $I$ stands for the unit operator.
The structure function (\ref{generalformofF}) can be expanded as
\begin{eqnarray}
    \Phi(J_3,J_0) = [J_0+J_3] [ I + J_0-J_3] \sum_{i=0}^{r} s_i (J_0+J_3-I)^i.
    \label{development of F(N)}
    \end{eqnarray}
In the last expression, the coefficients $s_i$ are given by
    \begin{eqnarray}
    s_0 = 1 \qquad s_i = \sum_{j_1 < j_2 < \cdots < j_i}  \kappa_{j_1} \kappa_{j_2} \cdots \kappa_{j_i} \quad
    (i = 1, 2, \cdots , r)
    \label{vieta formula}
    \end{eqnarray}
where the indices $j_1, j_2, \cdots , j_i$ take the values $1, 2, \cdots , r$. Reporting (\ref{development of F(N)}) in (\ref{G as difference of two F}),
one verifies
    \begin{eqnarray}
    G(J_3,J_0) = 2J_3 + {\Delta}_{\{\kappa\}}
    \label{development of G(N)}
    \end{eqnarray}
where the finite difference operator ${\Delta}_{\{\kappa\}}$ takes the form
\begin{eqnarray}
    {\Delta}_{\{\kappa\}} = [J_0+J_3][ I + J_0-J_3] \sum_{i=0}^{r} s_i (J_0+J_3-I)^i - [J_0+J_3+I][ J_0-J_3] \sum_{i=0}^{r} s_i (J_0+J_3)^i.
    \label{Delta}
    \end{eqnarray}
In the limiting case $\{\kappa\} \equiv \{0, 0, \cdots, 0\}$,  one verifies ${\Delta}_{\{\kappa\}}  = 0$  and hence the structure relations
(\ref{algebre}) reduces to usual  ones and the ordinary $u(2)$ algebra is recovered.

\subsection{Hilbert space}
\noindent Similar to standard $u(2)$ case, the unitary representations can be determined by considering a basis $\{\vert j,m \rangle, m = -j, -j+1, \cdots, j-1,j\}$, characterized by
an integer or half integer $j$, in which both Cartan subalgebra generators
$J_0$ and $J_3$ are diagonal
\begin{eqnarray}
 J_0 \vert j , m \rangle = j \vert j , m \rangle \qquad J_3 \vert j , m \rangle = m \vert j , m \rangle.
\end{eqnarray}
In this basis, the structure function acts as
\begin{eqnarray}
\Phi(J_3,J_0)  \vert j , m \rangle =  \Phi(j,m) \vert j , m \rangle.
\end{eqnarray}
with
\begin{eqnarray}
\Phi(j,m) = [j+m] [ 1 + j-m]
             [ 1 + \kappa_1(j+m-1)] [ 1 + \kappa_2(j+m-1)]\cdots
             [ 1 + \kappa_r(j+m-1)].
\end{eqnarray}
The action of raising and lowering generators write as follows
\begin{eqnarray}
 J_+ \vert j , m \rangle = \sqrt{\Phi(j,m)} \vert j , m + 1 \rangle \qquad J_- \vert j , m \rangle = \sqrt{\Phi(j,m+1)} \vert j , m - 1\rangle.
\end{eqnarray}
It is worth to note that the structure function is the product of the operators $J_+$ and $J_-$ (\ref{Phi(J_3,J_0}) and therefore the
eigenvalues $\Phi(j,m)$ should be non negative. Thus, the following condition
\begin{eqnarray}
[j+m] [ 1 + j-m]
             [ 1 + \kappa_1(j+m-1)] [ 1 + \kappa_2(j+m-1)]\cdots
             [ 1 + \kappa_r(j+m-1)] \geq 0 \label{conditionsurm}
\end{eqnarray}
must be fulfilled. Clearly, the positivity of the product in the left-hand side of this inequality depends on the parameters $\kappa_1, \kappa_2, \cdots, \kappa_r$.
This determines the dimension of the representations space of nonlinear $u(2)$ algebras defined through the structure function (\ref{Phi(J_3,J_0}). Obviuosly,
when  all the parameters $\kappa_i$ $(i = 1,2,\cdots,r)$ are in $\mathbb{R}_+$, the condition (\ref{conditionsurm}) is satisfied. In this case,  the algebra ${{u}}_{\{\kappa\}}(2)$ admits
a finite dimensional representations characterized by integer or half integer $j$ . The dimension is exactly $2j+1$  $(m = -j, -j+1. \cdots, j)$ as for the ordinary $u(2)$ algebra. The
situation changes when one or more parameters  are negative. Indeed, suppose that all parameters are positive except one, say $\kappa_i$ $( 1\leq i\leq r)$.
In this situation, the condition (\ref{conditionsurm}) gives
\begin{eqnarray}
m = -j, -j+1, \cdots, -j+E(-\frac{1}{\kappa_i})
\end{eqnarray}
where the symbol $E(x)$ stands for the integer part of $x$. To simplify, we assume that $-1/\kappa_i \in \mathbb{N}^{\star}$. Accordingly, the dimension of the Hilbert space is
$$ d = {\rm inf}(2j+1, d_i)$$
with
$$ d_i = 1 - \frac{1}{\kappa_i}.$$
In this respect for $\kappa_i < -\frac{1}{2j}$, the $(2j+1)$-dimensional representation space is truncated from $(2j+1)$ to $d_i$.
 This dimensional truncation  can be traduced by  the following nilpotent properties of the operators $J_+$ and $J_-$
$$ \big( J_- \big)^{d_i} =  \big( J_+ \big)^{d_i} = 0 $$
which differ from the usual case. Here, we have deliberately focused on the
case where only one parameter is non positive. However, we stress  that  these analysis can be repeated similarly to determine
the representation space dimension of ${{u}}_{\{\kappa\}}(2)$ involving two, three or more negative parameters.

\subsection{Contraction and ${\cal A}_{\{\kappa\}}$ algebra}
\noindent In order to establish the correspondence between the polynomial algebra ${{u}}_{\{\kappa\}}(2)$ and the generalized
Weyl-Heisenberg ${\cal A}_{\{\kappa\}}$  algebra
introduced in \cite{daoud3}, we review the Schwinger map in realizing  the ordinary algebra ${u}(2)$. In this realization,  the
generators  are defined  by means of two commuting pairs of ordinary (un-deformed or linear) bosons, say $\{ b_+^+ ,b_+^- \}$ and $\{ b_-^+ ,b_-^- \}$
acting on a  two-particle Fock space
${\cal F} = \{ \vert {n_+},{n_-} \rangle : n_+ \in \mathbb{N} , \
                      n_- \in \mathbb{ N } \}. $
The passage from boson state vectors
$\vert {n_+},{n_-} \rangle$  to angular momentum state vectors $\vert {j}, {m} \rangle$
is achieved  via the relations
\begin{equation}
 j := \frac{1}{2}(n_+ + n_-), \quad \qquad
 m := \frac{1}{2}(n_+ - n_-)
\label{eq:sch1}
\end{equation}
and
\begin{equation}
 \vert{j}, {m}\rangle \  \equiv \ \vert {j+m},{j-m}\rangle  \  = \  \vert {n_+},{n_-} \rangle.
\label{eq:sch2}
\end{equation}
In this picture, the actions of boson operators may thus be rewritten as
 \begin{eqnarray}
{b}_{\pm}^+ \; \;  \vert {j},{m}\rangle & = &
\sqrt{ {j{\pm}m+\frac{1}{2}+\frac{1}{2}}}\rangle\; \;
\vert{j+\frac{1}{2}}, {m{\pm} \frac{1}{2}}\rangle ,   \nonumber \\
{b}_{\pm}^-   \; \;  \vert{j},{m}\rangle & = &
\sqrt{ {j{\pm}m+\frac{1}{2}-\frac{1}{2}}}\rangle \; \;
\vert{j-\frac{1}{2}}, {m{\mp} \frac{1}{2}} \rangle,
\label{eq:qpbo2}
\end{eqnarray}
so that the bosons behave as ladder operators for the
quantum numbers $j$ and $m$ (with $\vert m \vert \le j$). In this realization, the four operators $J_\alpha$ ($\alpha = 0,3,+,-$)
are expressed as
\begin{equation}
J_0 \  := \ {{1}\over{2}}(N_+ + N_-) , \quad
J_3 \  := \ {{1}\over{2}}(N_+ - N_-) , \quad
J_+ \  := \  {b}_+^+ {b}_-^-     , \quad
J_- \  := \  {b}_-^+ {b}_+^-
\label{eq:gener1}
\end{equation}
in terms of usual creation, annihilation and number operators $(N_{\pm} = b_{\pm}^+b_{\pm}^-)$. They satisfy
the commutation rules
 \begin{equation}
[J_3,J_\pm   ] \ = \ \pm J_\pm ,  \qquad
[J_+,J_-     ] \ = \ {2J_3} ,  \qquad
[J_0,J_\alpha] \ = \ 0.
\label{eq: comm}
 \end{equation}
By restricting the Fock space ${\cal F}$ to its subspace generated by the vectors  such that  $n_+ + n_- = 2j$, one
verifies the identities
 \begin{equation}
J_0+J_3 = N ~~\qquad ~~J_0 -J_3 = 2j I - N,
 \end{equation}
where we replaced the number operator  $ N_+$ by $N$. Consequently,
 the structure function (\ref{generalformofF}) takes now the form
\begin{eqnarray}
    \Phi(J_3,J_0)\equiv \Phi_j(N)  = N ( (2j+1)I - N)
             ( I + \kappa_1 (N-I)) ( I + \kappa_2 (N-I))\cdots
             ( I + \kappa_r (N-I)).
    \label{generalformofF-WHA}
    \end{eqnarray}
 Mimicking the contraction procedure to pass from $su(2)$ algebra to harmonic oscillator algebra, we introduce the operators
\begin{eqnarray}
    a_+ = \frac{J_+}{\sqrt{2j}}~~ \qquad~~ a_- =  \frac{J_-}{\sqrt{2j}},
    \end{eqnarray}
and considering the situation where $j$ is large, one obtains
\begin{eqnarray}
    a_+ a_- = \Phi_{\infty}(N)  = N  ( I + \kappa_1 (N-I)) ( I + \kappa_2 (N-I))\cdots
             ( I + \kappa_r (N-I)).
    \end{eqnarray}
This is exactly the structure function  $F(N) \equiv \Phi_{\infty}(N) $  defining  the generalized Weyl-Heisenberg  ${\cal A}_{\{\kappa\}}$ introduced in \cite{daoud3}.
As a particular case, for $r=1$ and $\kappa_1 = \kappa$, the algebra ${\cal A}_{\{\kappa\}}$ reduces  to the generalized Weyl-Heisenberg algebra ${\cal A}_{\kappa}$ \cite{daoud1} . The
 commutation relations reduce to
\begin{eqnarray}\label{rel-A-1kappa}
[a^- , a^+] = I + 2 \kappa N \qquad [N , a^{\pm}] = \pm a^{\pm}.
    \end{eqnarray}
The corresponding structure function is now quadratic with respect to number operator \cite{daoud1}
\begin{eqnarray}\label{F1kappa}
    F(N)\equiv a^+a^- = N  ( I + \kappa (N-I)).
    \end{eqnarray}
To close this section, it is interesting to note that
Schwinger realizations of non linear $su(2)$ algebras were studied for some specific nonlinear scheme in \cite{Sunil1,Sunil2}. In this sense, it is natural
to ask: What kind of nonlinear $u(2)$ algebra can be realized in the Schwinger sense in terms of a pair of two generalized  bosons of type  ${\cal A}_{\kappa}$ (\ref{rel-A-1kappa})?
This issue constitutes the main of the following section.

\section{ Two ${\cal A}_{\kappa}$-boson realization of polynomial $u(2)$ algebras}
\subsection{ The realization}
\noindent In this section, we extend the usual Schwinger realization by replacing ordinary bosons by objects satisfying
the commutation rules of  ${\cal A}_{\kappa}$ algebra. In this order,  we consider two pairs of mutually commuting generalized Weyl-Heisenberg
algebra ${\cal A}_{\kappa}$ (\ref{rel-A-1kappa}). Each is  spanned by
 three linear operators $a_i^-$, $a_i^+$ and $N_i$ $(i = 1,2)$ satisfying the relations
\begin{eqnarray}
[a_i^- , a_i^+] = I + 2 \kappa N_i \qquad [N_i , a_i^{\pm}] = \pm a_i^{\pm}
\qquad \left( a_i^- \right)^{\dagger} = a_i^+ \qquad N_i^{\dagger} = N_i.
\label{thealgebra}
\end{eqnarray}
 We denote the corresponding Fock space, finite or infinite dimensional, by
\begin{eqnarray}
\mathcal{F}_{\kappa} = \mathcal{F}_{1} \otimes \mathcal{F}_{2} = \{ \vert
n_1 , n_2 \rangle, ~ n_1, n_2 ~~\mathrm{ranging} \}.
\end{eqnarray}
The finiteness of $\mathcal{F}_{\kappa}$ depends  on the value of the parameter $\kappa$. The actions of creation, annihilation and number operators on $\mathcal{F}_{\kappa} $
are defined by
\begin{eqnarray}
& & a_i^+ \vert n_1 , n_2 \rangle = \sqrt{F(n_i+1)} ~ \vert n_1 + s^-_i ,
n_2 + s^+_i \rangle,  \nonumber \\
& & a_i^- \vert n_1 , n_2 \rangle = \sqrt{F(n_i)} ~ \vert n_1 - s^-_i , n_2
- s^+_i \rangle,  \label{action  n} \\
& & a_1^- \vert 0, n_2 \rangle = 0, \quad a_2^- \vert n_1, 0 \rangle = 0,
\quad N_i \vert n_i \rangle = n_i ~\vert n_i \rangle,  \nonumber
\end{eqnarray}
where the quantity $s^{\pm}_i$ is defined by
\[
s^{\pm}_i = \frac{1}{2} ( 1 \pm (-)^i).
\]
In (\ref{action  n}), the quantities
\begin{eqnarray}\label{F(n)}
F(n_i) = n_i [1 + \kappa (n_i - 1)] \qquad i = 1, 2.
\end{eqnarray}
are the eigenvalues of the operators $F(N_i)$ given by (\ref{F1kappa}) \cite{daoud1}. They are
 quadratic  with respect to quantum numbers $n_i$ except the special case $\kappa = 0$ where
usual harmonic oscillators are recovered. In the Schwinger picture,  we realize the $u(2)$ operators  as
\begin{equation}
j_+ =a_1^{+} a_2^- , ~~~ j_- = a_1^- a_2^{+},~~~ j_3 = \frac{1}{2} ( N_1 - N_2 ) , ~~~ j_0 = \frac{1}{2} ( N_1 + N_2 ).
\end{equation}
Using the expression of the structure function $F(N)$ (\ref{F1kappa}) and the relations (\ref{thealgebra}), it is simply verified that
\begin{equation}
[ j_+ , j_- ] = 2j_3 \big( I - \kappa + 2\kappa j_0 (I + \kappa j_0)\big)
- 4\kappa^2 j_3^3 \qquad [j_3, j_{\pm}] = \pm j_{\pm} \qquad
    [j_0, j_{\alpha}] = 0 \label{su2-schwinger}
\end{equation}
with $\alpha = 0,3,+,-$.  When $\kappa = 0$, equation (\ref{su2-schwinger}) goes back to the
commutation relations satisfied by the angular momentum algebra. It is remarkable that the obtained algebra
is exactly  the cubic Higgs algebra \cite{2} (see also \cite{Debergh}) introduced to establish additional dynamical symmetries for
isotropic oscillator and Kepler problem.  This shows the relevance  the generalized Weyl-Heisenberg algebra ${\cal A}_{\kappa}$ introduced
in \cite{daoud1} in treating such kind of integrable systems.
\subsection{ Fock space}
\noindent Now, we examine the Hilbert space  of the nonlinear algebra satisfying the relations (\ref{su2-schwinger}).
Making use of parallel treatment of angular momentum, it is simple to obtain the unitary representations. Using the actions
of the creation and annihilation operators (\ref{action  n}), one gets
\begin{equation}\label{actions des j}
j_+ \vert n_1,n_2 \rangle = \sqrt{F(n_1+1)F(n_2)} \vert n_1+1,n_2-1 \rangle , ~~~ j_- \vert n_1,n_2 \rangle = \sqrt{F(n_1)F(n_2+1)}\vert n_1-1,n_2+1 \rangle
\end{equation}
where the function $F(n)$ should be positive. Henceforth, $\kappa \geq 0$ guarantees the positivity of $F(n)$ for any integer $n \in \mathbb{N}$.
However, for $\kappa < 0$, the function $F(n)$ is positive only when $n = 0, 1, \cdots, 1-1/\kappa$. For simplicity, we shall assume that $-1/\kappa \in \mathbb{N}^{\star}$.
Subsequently, for $\kappa \geq 0$, the Fock
space $\mathcal{F}$ is infinite dimensional:
\[
\mathcal{F}_{+}  \equiv \mathcal{F}_{\kappa \geq 0} = \{ \vert n_1 , n_2
\rangle, ~ n_1 \in \mathbb{N},~ n_2 \in \mathbb{N} \}.
\]
For $\kappa < 0$, the admissible values of the integers $n_1$ and $n_2$ are $0,1,\cdots, d-1$ with $ d = 1-1/\kappa$.  In this
case, the corresponding Fock space is
\[
\mathcal{F}_{-}  \equiv \mathcal{F}_{\kappa < 0}  = \{ \vert n_1 , n_2
\rangle, ~ n_1= 0, 1, \cdots, d-1; ~~n_2 =0, 1, \cdots, d-1 \},
\]
and therefore its dimension is $d^2$. Here again, to pass from the Fock number states to angular momentum basis, we introduce the quantum numbers $j$ and
$m$ defined as
\[
j = \frac{1}{2} (n_1 + n_2) ~~ \qquad ~~ m= \frac{1}{2} (n_1 - n_2),
\]
and we set the following correspondence
\[
\vert n_1 , n_2 \rangle \equiv \vert \frac{n_1 + n_2}{2} , \frac{n_1 - n_2}{2
} \rangle = \vert j , m\rangle.
\]
With this mapping, the Fock spaces $\mathcal{F}_{+} $ and $
\mathcal{F}_{-} $ decompose as
\[
\mathcal{F}_{+} = \bigoplus_{j=0}^{\infty}~~ \mathcal{F}_{j} =
\bigoplus_{j=0}^{\infty}~~ \{ \vert j , m \rangle , ~m = -j , -j + 1, \cdots
, j-1, +j\}
\]
and
\[
\mathcal{F}_{-} = \bigoplus_{j=0}^{d}~~ \mathcal{F}_{j} =
\bigoplus_{j=0}^{d-1} ~~\{ \vert j , m \rangle ,~ m = -j , -j + 1, \cdots ,
j-1, +j\}.
\]
Clearly, for $\kappa < 0$, the Schwinger realization produces only
angular momentum with $j = 0,{\frac{ 1}{2}}, 1, \cdots d-1$ so

that the the dimension of representations space $\mathcal{F}_{-} $ is $d^2$ ($\mathrm{dim} \mathcal{F}_{-} = d^2 $). This constitutes the
main difference with the case where $\kappa \geq 0$ where the full set of
$su(2)$ unitary representations are recovered.

\subsection{Fock-Bargmann realization and Coherent states}
Fock-Bargmann representation, based on coherent states formalism,  is a powerful method for obtaining closed analytic expressions for
various properties in quantum systems. We notice that coherent states for some quantum systems
described by nonlinear $su(2)$ and its noncompact counterpart $su(1,1)$ algebra are derived in \cite{Sadiq1,Sadiq2,Jagan,Cannata}.  For
the Higgs algebra defined by (\ref{su2-schwinger}), we shall construct a suitable analytical representation for the states belonging
to the subspace $\mathcal{F}_{j}= \{ \vert j , m \rangle , ~m = -j , -j + 1, \cdots
, j-1, +j\}$. To simplify our notations, we set
\[
\vert j , m \rangle = \vert n \rangle \quad \mathrm{with} \quad n = j+m .
\]
From (\ref{actions des j}), the actions of the generalized $su(2)$ algebra operators
rewrites
\begin{equation}\label{jacrtion surn}
j_- \vert n \rangle = \sqrt{f(n)} \vert n-1 \rangle
\qquad
j_- \vert n \rangle = \sqrt{f(n+1)} \vert n+1 \rangle
\qquad
j_3 \vert n \rangle = (n-j)\vert n+1 \rangle
\end{equation}
where the new function $f(n)$ is defined by
\begin{equation}\label{f(n)schwinger}
f(n) = F(n)F(2j-n+1)= n(2j+1-n)(1+\kappa(n-1))(1+\kappa(2j-n))
\end{equation}
in terms of the structure function of the generalized
Weyl-Heisenberg algebra ${\cal A}_{\kappa}$ given by
(\ref{F(n)}). For a representation characterized by the spin $j$, the coherent states are of the
form
\begin{equation}
\vert z \rangle = \sum_{n = 0}^{2j} a_n z^n \vert n \rangle
\end{equation}
where $z$ is a complex variable, $n = j+m$ and the $a_n$ coefficients to be
determined. In the Bargmann realization any vector is realized as follows
\begin{equation}\label{Fock-bargmann}
\vert n \rangle \longrightarrow a_{n} z^{n} \equiv \langle \bar z\vert n
\rangle
\end{equation}
and the operator $j_-$ is assumed acts as a derivation with respect to the complex variable $z$
\begin{equation}\label{actionjmoins}
j_-\longrightarrow \frac{d}{dz}.
\end{equation}
Thus, using the expression of the action of $j_-$ (\ref{jacrtion surn}) and the
Bargamnn correspondence (\ref{Fock-bargmann}), one shows that the coefficients $a_n$
satisfy the recursion relation
\begin{equation}
n a_{n}= \sqrt{f(n)}a_{n-1}
\end{equation}
where $f(n)$ is given by (\ref{f(n)schwinger}). This yields
\begin{equation}
a_{n}= a_0 \frac{\sqrt{f(n)!}}{n!}
\end{equation}
where $f(n)!= f(1) f(2) \cdots f(n)$ with $f(0)! = 1$. The coefficient $a_o$
is obtained by imposing  the normalization condition for the states $\vert z \rangle$ . Finally, one gets
\begin{equation}
\vert z \rangle = \mathcal{N}^{-1} \sum_{n=0}^{2j} \frac{\sqrt{f(n)!}}{n!}
z^n \vert n \rangle
\end{equation}
where the $\mathcal{N}$ normalization factor is given by the sum
\begin{equation}
|\mathcal{N}|^2 = \sum_{n = 0}^{2j} \frac{f(n)!}{(n!)^2} \vert z \vert^{2n}.
\end{equation}
 In the limiting case $\kappa
\rightarrow 0$, one recovers the $su(2)$ coherent states and the
standard Bargamnn realization based on spin coherent states.
In this realization, the operators $j_3$ and $j_+$ act,
respectively, as follows
\begin{equation}\label{actionjplus}
j_3\longrightarrow {z} \frac{d}{d{z}} - j \qquad j_+\longrightarrow {z} %
\bigg( 1 + \kappa{z} \frac{d}{d{z}}\bigg) \bigg( 2j - {z} \frac{d}{d{z}}%
\bigg) \bigg( 1 + (2j-1)\kappa - \kappa {z} \frac{d}{d{z}}\bigg).
\end{equation}
Any state $\vert \Psi \rangle$, belonging to angular momentum  space ${\cal F}_j$,
\[
\vert \Psi \rangle = \sum_{n=0}^{2j} \Psi_{n}\vert n \rangle
\]
is represented by  an analytical function as follows
\begin{equation}  \label{psi(z)}
\vert \Psi \rangle \longrightarrow \Psi(z) = \mathcal{N} \langle \bar z
\vert \Psi\rangle = \sum_{n=0}^{2j} \Psi_{n} f_{n}(z)
\end{equation}
where the monomials $f_{n}(z)$ are the analytical functions associated with
the vectors $n$
\begin{equation}
f_{n} (z) = \mathcal{N} \langle \bar z \vert 2n \rangle = \frac{\sqrt{f(n)!}
}{n!} z ^{n}.
\end{equation}
Using the realization (\ref{actionjmoins}) and the results (\ref{actionjplus}), one gets
\begin{equation}
j_3 f_{n}(z) = ( n- j) f_{n}(z),
\qquad
j_+ f_{n}(z) = \sqrt{f(n+1)} f_{n+1}(z),
\qquad
j_- f_{n}(z) = \sqrt{f(n)} f_{n-1}(z),
\end{equation}
to be compared with (\ref{jacrtion surn}).

\section{ Concluding remarks}

\noindent To summarize, we introduced a new  variant  of polynomial $u(2)$
algebras. The corresponding unitary representations were constructed. We have shown that the dimension of the corresponding Hilbert
space can be modified for some specific values of the parameters $\kappa_1, \kappa_2, \cdots, \kappa_r$. Via a contraction
procedure to the multi-parametric polynomial Weyl-Heisenberg algebra ${\cal A}_{\{\kappa\}}$
introduced in \cite{daoud3} is obtained. The second facet of this letter concerned the realization of Higgs algebra by means
of a pair of nonlinear (quadratic) boson generating  the one parameter algebra ${\cal A}_{\kappa}$. As a further prolongation,
it would be interesting to extend the nonlinear bosonic Schwinger realization for other Lie algebras.
Another interesting issue concerns other polynomial $su(2)$ extensions as for instance one defined by the following commutation relations
$$
[J_+, J_-] = P(J_3) \ , \quad [J_3, J_+] = G(J_3) \, J_+ \ , \quad [J_3, J_-] = - J_- \ G(J_3)
$$
where $P$ and $G$ are polynomials of the diagonal operator $J_3$.

\end{document}